# Surface pattern determined by vertical convection on Rayleigh-Taylor instability

Michiko Shimokawa

*Center for Frontier Science, Chiba University*

*1-33, Yayoi-cho, Inage-ku, Chiba-shi, Chiba, 263-8522, Japan*

Abstract

Relationship between a surface pattern and vertical convections is studied in a condition of Rayleigh-Taylor instability. The vertical convections change with the case configuration and the aspect ratio $r/h$ of the case, where $r$ and $h$ show radius of case and height of the lighter solution. Fractal pattern is observed at $r < h$, and cell pattern is formed at $r > h$. According to the changes of the surface pattern, the vertical convection transfers to some convections from coupled convections. This result shows that the surface pattern is strongly influenced by the vertical convection.

Keywords: Pattern formation, Convection, Experiment, Rayleigh-Taylor instability

Email: M. Shimokawa (shimokawa@physics.s.chiba-u.ac.jp)

Tel: +81-43-290-3248

Fax: +81-43-290-3523

1. **INTRODUCTION**

Convection is found in various field of science, such as the mantle[1-3], atmospheric circulation[4-5], thermohaline circulation[6] and bioconvection[7-9]. Especially, Rayleigh-Benard convection, which is representative example as heat transport, has absorbed topics in physics, since it has deep relationship with chaos observed in the process to turbulence flow from laminar flow[10-17]. In the process to turbulence flow from laminar flow, Benard convection appears. Various patterns, such as the hexagonal and tetragonal shapes and belt-like pattern, are observed at the surface of convection. The pattern and its size depend on configuration of experimental container, and then the vertical flow changes. It is reported that the surface pattern is determined by the vertical convection[16−17].

Recently, a new type of annihilative fractal formed in the condition of Rayleigh-Taylor instability was reported [18]. In the formation process of the fractal, a heavier solution sinks toward the bottom and a lighter solution rises toward the surface according to volume conservation. This behavior causes the vertical convection. The surface pattern is determined by the vertical flow, which is similar to Benard convection.

In this paper, I report on the dependence of surface pattern by vertical flow in experiments with some configurations of cases and aspect ratio.

2. EXPERIMENTAL PROCEDURE

Rayleigh-Taylor instability, [19-21] which shows vertical disturbances at the surface, occurs when heavier solution is put on lighter solution. Milk and glycerine solution are used as lighter solution. The densities of milk and glycerine solution are 1.03g/ml and 1.21 g/ml, respectively. The glycerin solution consists of glycerin (WAKO), white watercolour and water. White watercolor 0.50g is melted in water 100g for clarity of the surface pattern. The solution and glycerine are mixed with the same masses, and the glycerine solution is completed. The magnetic fluid with density 1.40 g/ml (Taiho Kozai, Ferri Colloid W-40) is used as heavier solution. This is a colloidal mixture composed of magnetite, but it is not magnetized. Thus, the magnetic fluid can be recognized just as a high density liquid in our experiments. The both densities of milk and glycerine solution are lighter than that of magnetic fluid, and it satisfies the condition of Rayleigh-Taylor instability. The colours of heavier and lighter solution are black and white, respectively, because of image binarization and clear visualization.

The configurations of the experimental cases are cylinder, star pole, triangle pole and doughnut-shaped case. The cross-section view of doughnut-shaped case is shown in Fig. 1 (b).

An experimental procedure is as follows: The lighter solution is poured into a tall beaker until it reaches to the mark of $h$ cm as shown in Fig. 1 (a). The beaker stands for 10 min or more for the fluid to come to rest. Next, magnetic fluid of 0.2 ml droplet is placed silently at the surface center of lighter solution with a pipette, as shown in Fig. 1 (a). The experiment is recorded using a digital video camera (Sony HDR-FX1). The shutter speed of the camera is $1/60$ s and the resolution is $640 \times 480$ pixels. The time evolutions of a surface pattern and a vertical flow are captured from the position of Camera 1 and Camera 2 in Fig. 1, respectively. A macro lens (Raynox DCR-150) is attached to the digital video camera to increase clarity.

3. RESULTS AND DISCUSSION

In order to confirm an effect of a configuration of an experimental case, surface pattern is investigated in experiments using a circular case in Fig. 2 (a), a star case in Fig. 2 (b) and a triangle case with a corrugated wall in Fig. 2 (c). In this experiment, milk is used as lighter solution. These analysed results by box counting method shows in Figs. 2 (d)-(e), which correspond to Figs. 2 (a)-(c), respectively. The results show that all patterns in Figs. 2 (a)-(c) have fractal structure, whose dimensions are 1.88, 1.88 and 1.89 respectively. These fractal dimensions are very close to each other. Therefore, these patterns are regarded as the same. These have a common point that the branches of fractal pattern gather around at the gravity center.

In order to investigate how pattern is formed when the solution cannot access the gravity center, the donuts shaped case in Fig. 1 (b) is used. The internal diameter and the width of the channel are 5.0 cm and 2.5 cm as shown in Fig. 1 (b), respectively. Figure 3 (a) shows the surface pattern observed in experiments using a donut case. The expanded picture is shown in Fig. 3 (b). The pattern is investigated by density-density correlation function and box counting method, whose results show in Figs. 3 (c) and (d). These results show that the pattern in Fig. 3 (b) is fractal, and the obtained dimensions are 1.82 and 1.83, respectively. Branches in Fig. 3 (b) gather at median line between internal and external walls. This pattern is different from ones in Figs. 2 (a)-(c) whose branches gather around at the gravity center of the container.

Above results imply a close relationship between the surface pattern and the vertical flow. The coupled convection occurs between a wall and the kitty-cornered wall, which is supported by reference [18]. In Fig. 2 the coupled convections are connected around at the gravity center. On the other hand, the coupled convections gather toward the medium line of walls in Fig. 3. These results lead the consideration that the convection is important for the determination of the surface pattern.

The surface pattern is also investigated using the case in varies of radius $r$ and depth $h$, because it is considered that the experimental case size is important according to the report of Benard convection. In the experiments, glycerine solution is used as lighter fluid instead of milk. It is reported that the dynamics of the surface pattern formation is similar to experiments using milk[18]. Figures 4 (a) and (b) show surface patterns in the conditions of $r = 5.3$ cm, $h = 7.0$ cm and $r = 5.3$ cm, $h = 4.0$ cm, respectively. A new type of cell pattern, shown in Fig. 4 (b), is also found in the experiments. When fractal and cell patterns are formed, the vertical flows in Figs. 4 (c) and (d) are observed. In the formation of fractal pattern, magnetic fluid gathers at the center of the beaker and sinks vertically as shown in Fig. 4 (c). On the other hand, in the formation of the cell pattern magnetic fluid sinks at several points; branches do not gathers at the center as shown in Fig. 4 (d).

Next, I investigate the transition between fractal and cell patterns in experiments with varieties of radius $r$ and depth $h$. As shown in Fig. 5 (a), the fractal pattern emerges at $r < h$, and the cell pattern is observed at $r > h$. This result shows that the pattern transition to fractal pattern from cell pattern occurs at $r = h$. At the same time, the type of the vertical convection transfers to that of Fig. 4 (c) from Fig. 4 (d).

Let us consider the relationship between the vertical convection and the surface pattern. When the magnetic fluid sinks, glycerine rises according to the volume conservation of two solutions. This behaviour causes the convection by the glycerine solution and magnetic fluid. In the process, glycerine solution, which rises toward the surface, displaces to the magnetic fluid, and sinking regions by magnetic fluid draw the line of the surface pattern.

If several convections develop as shown in Fig. 5 (b), then magnetic fluid at the surface is displaced by rising of the glycerine at several regions. It leads the formation of cell pattern as shown in Fig. 4 (b), whose line is drawn at the boundary of convections. On the other hand, only coupled convections exist between walls in the formation of fractal as shown in Fig. 5 (c). The sinking point, shown as heavy line in Fig. 5 (c), is only at the center, where almost agrees with the gravity center. According to the flow, branches of fractal pattern also gather at the gravity center. This is why pattern has a symmetrical geometric for the gravity center. This consideration shows the vertical flow, especially the position of sinking point, is important for the determination of the surface pattern.

The result of Fig. 3 also supports the consideration. Magnetic fluid sinks at the center line between walls, and the coupled convection develops along the line. The surface pattern depends on the convection, and branches gather at the center line in Fig. 3 (b). These results show that it is important for the fractal formation that magnetic fluid flows toward only one region at the surface.

It is considered that the formation process of the fractal is similar to viscous fingering[23-26], that the solution of low viscosity breaks in that of high viscosity. In this experiment, solutions of lower and higher viscosity correspond to glycerine and magnetic fluid, respectively. Previous report[18], that fractal pattern starts to form from surrounding of pattern. This result supports the above consideration. Number of sinking point as singularity, however, is difference between viscous fingering and this experiments. Ordinary viscous fingering has one singularity. On the other hand, some singularity points exist in this experiment. In future, it would be interesting to compare wavelengths between this fractal structure and ordinary viscous fingering [27-28]. In addition, it would be interesting to investigate the pattern transition in the variety of the viscosity.

4. SUMARRY

I investigate the surface pattern at the variety of case configuration and aspect ratio. In this experiment, it is clarified that the surface pattern depends on the vertical flow. In the formation of fractal pattern, magnetic fluid at the surface gathers at the center and sinks according to coupled convection. On the other hand, magnetic fluid sinks at some regions in the formation of the cell pattern. The transition of surface pattern and vertical flow occurs at $r = h$.

References

1. Taras Gerya, Future directions in subduction modelling, *Journal of Geodynamics* **52** (2011) 344-378.

2. A. W. Hofmann, Mantle geochemistry: the message from oceanic volcanism, *Nature* **385** (1997) 219-229.

3. R. D. van der Hilst, S. Widiyantoro and E. R. Engdahl, Evidence for deep mantle circulation from global tomography, *Nature* **386** (1997) 578-584.

4. B. E. Mapes, Gregarious tropical convection, *Journal of the atmospheric Sciences* **50** (1993) 2026-2037.

5. C. Zhang, Madden-Julian Oscillation, *Review of Geophysics* **43** (2005) 1-36.

6. E. Sayin, C. Eronat, S. Uckac and S. T. Besiktepe. Hydrography of the eastern part of the Aegean sea during the eastern Mediterranean transient (EMT), *Journal of Marine Systems* **88** (2011) 502-515.

7. T. J. Pedley, N. A. Hill and J. O. Kessler, The growth of bioconvection patterns in a uniform suspension of gyrotactic micro-organisms, *Journal of Fluid Mechanics* **195** (1988) 223-237.

8. C. R. Williams and M. A. Bees, A tale of three taxes: photo-gyro-gravitactic bioconvection, J. Exp. Biol. **214** (2011) 2398-2408.

9. A. V. Kuznetsov, Nanofluid bioconvection: interaction of microorganisms oxytactic upswimming, nanoparticle distribution, and heating/cooling from below, *Theor. Comput. Fluid Dyn.* **24** (2011) ?.

10. T. H. Solomon and J. P. Gollub, Chaotic particle transport in time-dependent Rayleigh-Benard convection, *Phys. Rev.* A **38** (1988) 6280-6286.

11. E. D. Siggia and A. Zippelius, Pattern Selection in Rayleigh-Benared Convection near thereshold, Phys. Rev. Lett. **47** (1981) 835-838.

12. M. Krichnan, V. M. Ugaz and M. A. Burns, PCR in a Rayleigh-Benard Convection Cell, Science **25** (2002) 793.

13. E. Brown, A. Nikolaenko and G. Ahlers, Reorientation of the Large-Scale Circulation in Turbulent Rayleigh-Benard Convection, Phys. Rev. Lett. **95** (2005) 084503.

14. M. Sano and Y. Sawada, Measurement of the Ltapunov Spectrum from a Chaotic Time Series, *Phys. Rev. Lett.* **55** (1985) 1082–1085.

15. S. Chandrasekhar, *Hydrodynamic and Hydro magnetic stability* (Oxford University Press, 1961).

16. P. Ball, *The Self-Made Tapestry* (Oxford University Press, 2001).

17. M. C. Cross and P. C. Hohenberg, Pattern formation outside of equilibrium, *Review of Modern Physics* **65** (1933) 85-1112.

18. M. Shimokawa and S. Ohta, Annihilative fractals formed in Rayleigh-Taylor instability, *Fractals* **19** (2011) 1-8.

19. D. H. Sharp, An overview of Rayleigh-Taylor instability, *Physica D* **12** (1984) 3-10.

20. D. L. Youngs, Three-dimensional numerical simulation of turbulent mixing by Rayleigh–Taylor instability, *Phys. Fluids A* **3** (1991) 1312-1321.

21. M. Sano and Y. Sawada, *Turbulence and Chaotic Phenomena in Fluid* (North-Holland Amsterdam 1984) 167.

22. S. W. Morris, E. Bodenschatz, D. S. Cannell and G. Ahlers, Spiral defect Chaos in large aspect ratio Rayleigh-Benard convection, *Phys. Rev. Lett.* **71** (1993) 2026-2029.

23. Y. Nagatsu *et. al*. Miscible viscous fingering involving viscosity changes of the displacing fluid by chemical reactions, *Phys. Fluids*, **22** (2010) 024101-1-024101-13.

24. X. Cheng *et. al*. Towards the zero-surface-tension limit in granular fingering instability, *Nature Physics* **4** (2008) 234-237.

25. D. N. Bankar *et. al.* Segregation of fractal aggregates grown from two seeds, *Phys. Rev. E,* **75** (2007) 051401-1-051401-5.

26. O. Praud and H. L. Swinney, Fractal dimension and unscreened angles measured for radial viscous fingering, *Phys. Rev. E*. **72** (2005) 011406-1-011406-10.

27. M. G. Moore, A. Juel, J. M. Burgess, W. D. McCormick and H. L. Swinney, Fluctuations Viscous fingering, *Phys. Rev. E* **65** (2002) 030601(R).

28. L. Paterson, Diggusion-Limited Aggregation and Two-Fluid Displacements in Porous Media, *Phys. Rev. Lett.* **52** (1984) 1621-1624.

29. L. E. Johns and R. Narayanan, The Rayleigh–Taylor instability of a surface of arbitrary cross section with pinned edges, *Physics of Fluids* **23** (2011) 012108-.012110.

Acknowledgements

I thank H. Honjo and S. Ohta for proposing the experiment with the magnetic fluid and for fruitful discussions. I am also grateful to K. Yamamoto, T. Takami, N. Mitarai and H. Nakanishi for helpful suggestions; M. Homma and A. Saeki for discussions. This study is supported.by Japan Society Promotion of Science.

Fig. 1

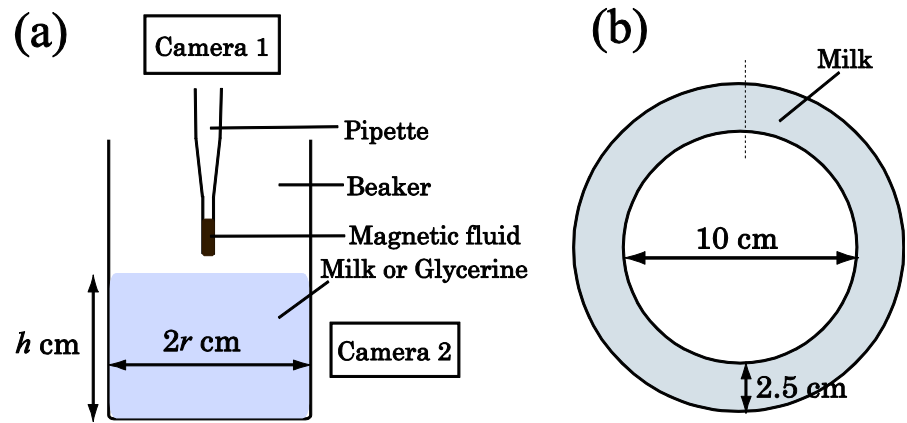

Figure 2

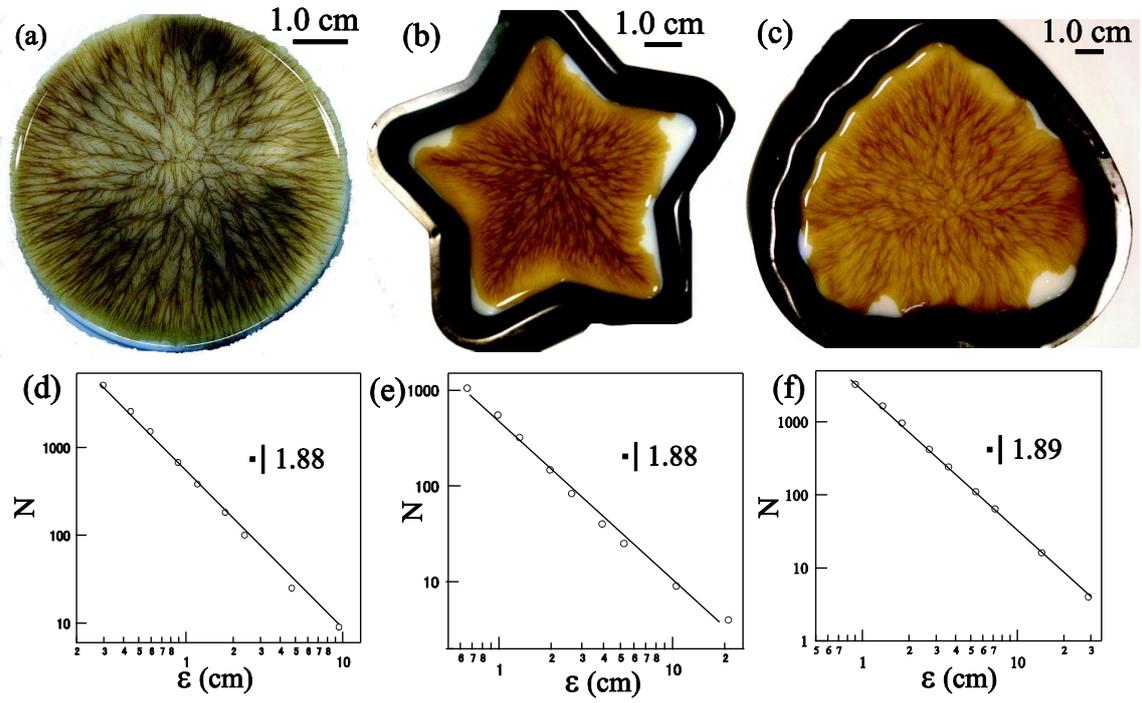

Figure 3

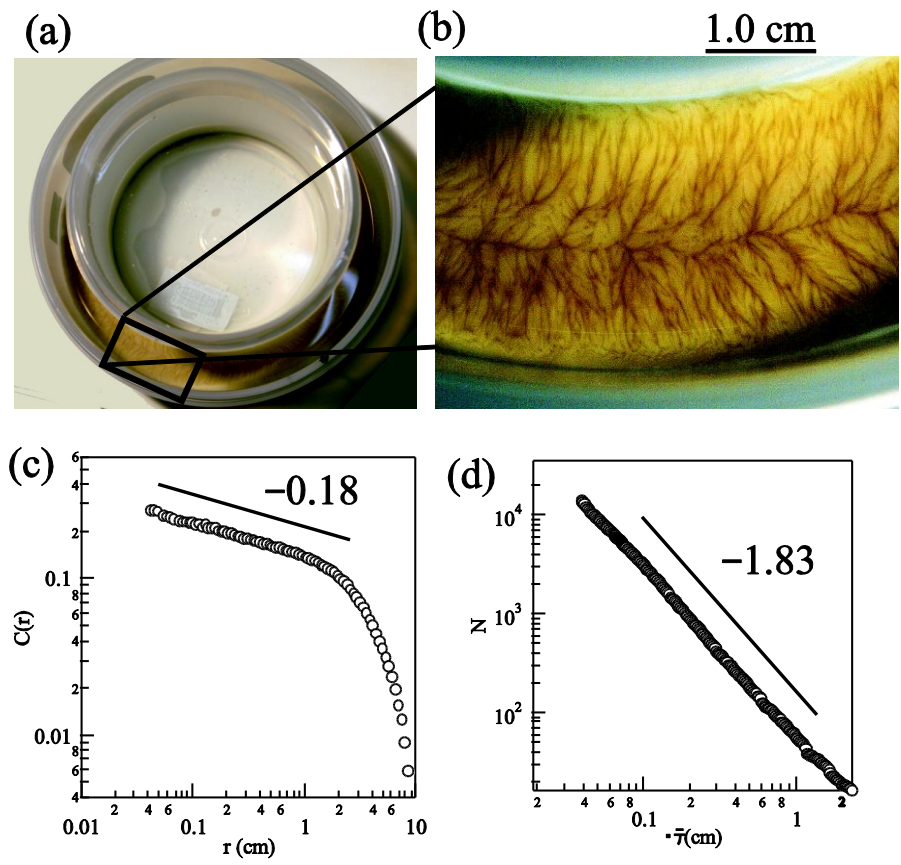

Figure 4

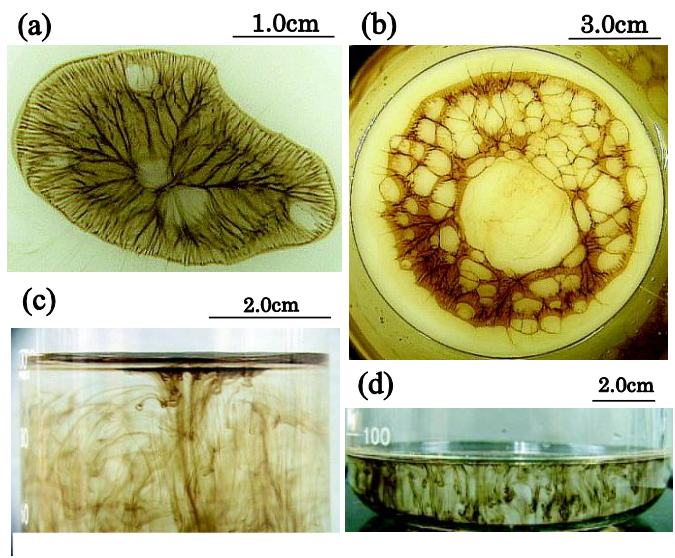

Figure 5

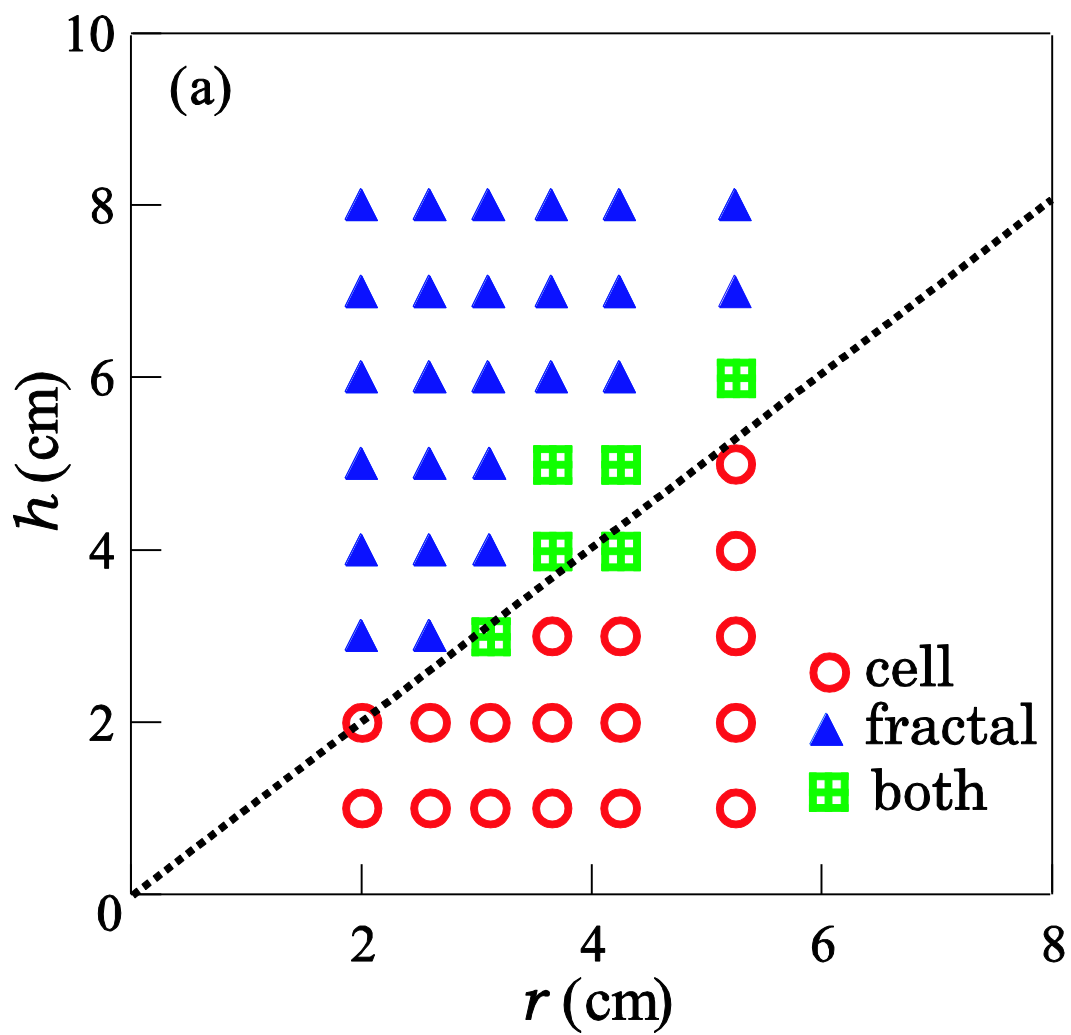
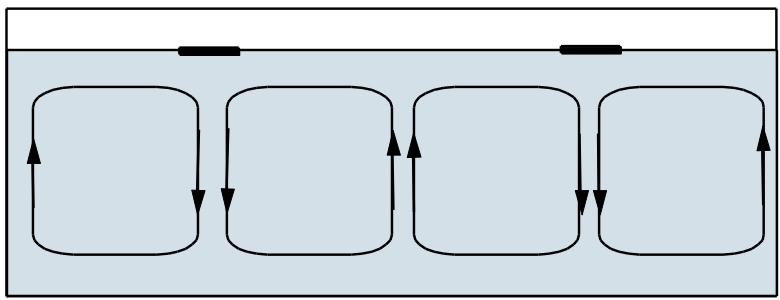
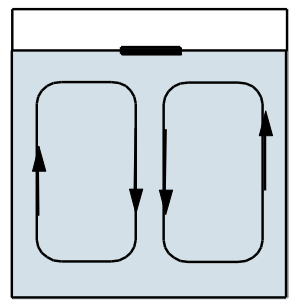

Figure captions

Fig. 1 (Color online) (a) Schematic drawing of experimental procedure. Radius of beaker and depth of glycerine solution are shown as $r$ and $h$. The surface pattern and vertical flow of magnetic fluid are captured at the positions of Camera 1 and 2, respectively. (b) Schematic drawing of donut-shaped case. The internal diameter and the width of channel are 5.0 cm and 2.5 cm, respectively. Milk as the lighter solution is poured into the channel.

Fig. 2 (Color online) (a)-(c) Fractal patterns obtained from experiments with (a) circle case, (b) star case and (c) triangle case, respectively. (d)-(f) Fractal dimensions obtained from box counting method. The data correspond to these of (a)-(c), respectively. Box sizes and box numbers represent as $\varepsilon$ and $N(\varepsilon)$.

Fig. 3 (Color online) (a) Snapshot of surface pattern obtained from experiments of doughnut-shaped container in Fig. 1(b). (b) Expanded picture of (a). (c)

Fractal dimensions obtained from density-density correlation function. The distance between two pixels and correlation function represent as $r$ and $C(r)$. (d) Fractal dimensions obtained from box counting method. Box sizes and box numbers represent as $\varepsilon$ and $N(\varepsilon)$.

Fig. 4 (Color online) (a), (b) Surface patterns captured by Camera 1 in Fig. 1. Fractal pattern of (a) is observed at $r = 3.0$ cm and $h = 6.0$ cm. Otherwise, cell pattern of (b) is observed at $r = 5.0$ cm and $h = 2.5$ cm. (c) and (d) are vertical flows, captured using Camera 2 of Fig. 1. In the observation of vertical flows, glycerine solution is used without white watercolour for visualization. The pictures are observed in the condition of (c) $r = 5.0$ cm, $h = 6.5$ cm and (d) $r = 5.0$ cm, $h = 2.0$ cm. These correspond to the observed region of fractal and cell patterns, respectively.

Fig. 5 (Color online) (a) Phase diagram on surface pattern. The radius of beaker and depth of glycerine are shown as $r$ and $h$. Open circles and closed triangles show cell and fractal patterns, respectively. The region, shown as opened squares, is where both of cell pattern and fractal pattern are observed.

(b), (c) Schematic drawings of vertical flow in the observation of cell and fractal patterns, respectively.